\documentclass{article}

\usepackage[english]{babel}

\usepackage[a4paper,top=2cm,bottom=2cm,left=3cm,right=3cm,marginparwidth=1.75cm]{geometry}

\usepackage{amsmath}
\usepackage{graphicx}
\usepackage{hyperref}
\hypersetup{colorlinks,allcolors=blue}

\usepackage{booktabs}

\usepackage{authblk}
\author[1]{Edoardo Loru}
\author[2]{Matteo Cinelli}
\author[3]{Maurizio Tesconi}
\author[2]{Walter Quattrociocchi}
\affil[1]{Department of Computer, Control and Management Engineering, Sapienza University of Rome}
\affil[2]{Department of Computer Science, Sapienza University of Rome}
\affil[3]{Institute of Informatics and Telematics, National Research Council}

\newcommand\alert[1]{\textcolor{red}{\textbf{#1}}}

\title{The influence of coordinated behavior on toxicity}
\date{}

\begin{document}
\maketitle

\begin{abstract}
In the intricate landscape of social media, genuine content dissemination may be altered by a number of threats.
Coordinated Behavior (CB), defined as orchestrated efforts by entities to deceive or mislead users about their identity and intentions, emerges as a tactic to exploit or manipulate online discourse. This study delves into the relationship between CB and toxic conversation on X (formerly known as Twitter). Using a dataset of 11 million tweets from 1 million users preceding the 2019 UK general election, we show that users displaying CB typically disseminate less harmful content, irrespective of political affiliation. However, distinct toxicity patterns emerge among different coordinated cohorts. Compared to their non-CB counterparts, CB participants show marginally higher toxicity levels only when considering their original posts. We further show the effects of CB-driven toxic content on non-CB users, gauging its impact based on political leanings. 
Our findings suggest that CB only has a limited impact on the toxicity of digital discourse.

\noindent
\alert{Please cite published version:} \href{https://doi.org/10.1016/j.osnem.2024.100289}{https://doi.org/10.1016/j.osnem.2024.100289}
\end{abstract}

\section{Introduction}
\label{intro}
Social media are nowadays one of the main arenas for public debate, where users get their information and interact with other peers under the potential influence of feed algorithms that are used to prioritize their engagement with like-minded content \cite{valensise2023drivers,del2016echo,huszar2022algorithmic}. According to recent studies, such systems can challenge democracy in various ways~\cite{gonzales2023science1, guess2023science2, guess2023science3}. Problems include the fast spread of false information~\cite{del2016spreading, lazer2018science, bovet2019influence, juul2021comparing, pierri2023propaganda}, more division among groups \cite{garret2009, bail2018exposure,tucker2018social,cinelli2021echo,falkenberg2022growing}, and harmful behaviors online~\cite{cheng2015antisocial, cinelli2021dynamics, johnson2023offlinehate}. Despite efforts to fix these issues, solutions are hard to find \cite{guess2023science2, guess2023science3}.

Further complicating this ecosystem is the phenomenon of Coordinated Behavior (CB), which can be defined as an unexpected, suspicious, or exceptional similarity among users of a group~\cite{nizzoli2021coordinated}. 
Social media campaigns, such as online activism, protests, and disinformation campaigns \cite{starbird2019disinformation, vargas2020detection, shu2020mining}, generally involve participants coordinating their actions to disseminate content widely. CB differentiates from Coordinated Inauthentic Behavior (CIB) which, according to Meta's definition, is ``the use of multiple Facebook or Instagram assets, working in concert to engage in inauthentic behavior, where the use of fake accounts is central to the operation". More in detail, according to the platform's Community Standards\footnote{{https://transparency.fb.com/en-gb/policies/community-standards/inauthentic-behavior/}}, the concept of inauthentic behavior refers to people who ``misrepresent themselves on Facebook, use fake accounts, artificially boost the popularity of content or engage in behaviors designed to enable other violations".

Initially, scientific research focused on the benefits of coordination for social movements. However, it has become evident that benign actors, such as activists, are using similar techniques, and malicious actors are engaged in political astroturfing~\cite{keller2020political} and the dissemination of inappropriate content. Coordinated behavior on social media can have negative consequences, including distorting public opinion and contributing to the polarization of society. Recognizing these problems and in the light of contrasting scientific results \cite{ruths2019misinformation,eady2023exposure}, researchers and practitioners are working on strategies to identify, characterize, and mitigate coordinated behavior~\cite{pacheco2021uncovering,cinelli2022coordinated,schoch2022coordination}.
In particular, the characterization of coordinated groups is a crucial aspect discussed in the existing literature \cite{pacheco2021uncovering}. This can be done at different levels of depth but remains essential due to the absence of ground-truth data for detection tasks.
To assess the harm of coordinated behavior, established methods in the literature can be used, primarily the analysis of content shared by coordinated users, which includes the identification of fake news \cite{kai2019defend, shaoo2021multiple}, hate speech and toxicity detection \cite{cinelli2021dynamics, Vigna2017HateMH}.

In the context of the rapidly evolving digital landscape, the 2019 UK general election provides a pertinent setting to explore the dynamics of online behavior. Our study aims to disentangle the complex relationship between CB and the prevalence of toxic content on the X platform (formerly called Twitter), which we view as a ``rude, disrespectful, or unreasonable comment that is likely to make people leave a discussion", as per Perspective API's definition \cite{perspectiveAPI}. For this purpose, we analyze a dataset encompassing 11 million tweets from a pool of 1 million users.
From our analysis, an interesting observation emerges: users exhibiting high coordination tend to disseminate less toxic content. This propensity holds regardless of their political leaning, suggesting that coordination might not necessarily be synonymous with malicious intent or negative discourse. However, distinct patterns of toxic behavior emerge when we delve deeper into the different communities of coordinated users, indicating a varied landscape of content sharing even within these coordinated groups. Furthermore, we observe that the extensive retweeting activity of coordinated users affects the toxicity levels they display, and that for coordinated and non-coordinated users original tweets have marginally higher toxicity levels than retweets.
Beyond these analyses, our study delves into the potential impact of content stemming from coordinated efforts. Specifically, we explore how interacting with toxic content, particularly when associated with CB, affects the behaviors of non-coordinated users. A key aspect of this analysis is assessing the potential role of political orientation in such interactions. Does a user's political leaning amplify or attenuate their reaction to CB-driven toxicity?
After carefully considering these dynamics, our results point towards CB only playing a minor role in the toxicity observed in online conversations.

\section{Materials and Methods}
\label{sec:mat_and_methods}

\subsection{Data}
\label{sec:data}

\begin{table}[t]
    \centering
    \caption{Data collected via hashtags \cite{nizzoli2021coordinated}. Neutral (N) hashtags have been assigned a political leaning score of $0$, whereas hashtags linked to the Labour (L) and Conservative (C) parties have been assigned a score of $-1$ and $+1$, respectively.}
    \begin{tabular}{lcrr} 
        \toprule
        Hashtag & Leaning & Users & Tweets\\
        \midrule
        \#GE2019               & N $(0)$ & 436,356 & 2,640,966 \\
        \#GeneralElection19    & N $(0)$ & 104,616 & 274,095 \\
        \#GeneralElection2019  & N $(0)$ & 240,712 & 783,805 \\
        \cmidrule(lr){1-1}
        \#VoteLabour           & L $(-1)$ & 201,774 & 917,936 \\
        \#VoteLabour2019       & L $(-1)$ & 55,703 & 265,899 \\
        \#ForTheMany           & L $(-1)$ & 17,859 & 35,621 \\
        \#ForTheManyNotTheFew  & L $(-1)$ & 22,966 & 40,116 \\
        \#ChangeIsComing       & L $(-1)$ & 8,170  & 13,381 \\
        \#RealChange           & L $(-1)$ & 78,285 & 274254 \\
        \cmidrule(lr){1-1}
        \#VoteConservative     & C $(+1)$ & 52,642 & 238,647 \\
        \#VoteConservative2019 & C $(+1)$ & 13,513 & 34,195 \\
        \#BackBoris            & C $(+1)$ & 36,725 & 157,434 \\
        \#GetBrexitDone        & C $(+1)$ & 46,429 & 168,911\\
        \midrule
        Total & & 668,312 & 4,983,499\\
        \bottomrule
    \end{tabular}
    \label{tab:data_hashtags}
\end{table}

\begin{table}[t]
\centering
\caption{Data collected via accounts \cite{nizzoli2021coordinated}. The two accounts linked to the Labour party (L) have been assigned a political leaning score of $-1$, whereas the two linked to the Conservative party (C) a score of $+1$.}
    \begin{tabular}{lcrrr}
        \toprule
        & \multicolumn{2}{c}{Production} & \multicolumn{2}{c}{Interactions}\\
        \cmidrule(lr){2-3} \cmidrule(lr){4-5}
        Account & Leaning & Tweets & Retweets & Replies\\
        \midrule
        @jeremycorbyn  & L $(-1)$ & 788    & 1,759,823 & 414,158\\
        @UKLabour      & L $(-1)$ & 1,002  &  325,219  &  79,932\\
        @BorisJohnson  & C $(+1)$ & 454    &  284,544  & 382,237\\
        @Conservatives & C $(+1)$ & 1,398  &  151,913  & 169,736\\
        \midrule
        Total & & 3,642 & 2,521,499 & 1,046,063\\
        \bottomrule
    \end{tabular}
    \label{tab:data_accounts}
\end{table}

\begin{table}[t]
    \centering
    \caption{Accounts excluded from the set of ``coordinated users'' studied in this work, and manually labeled as ``non-coordinated''.}
    \begin{tabular}{ll}
        \toprule
        Account & Description\\
        \midrule
        @jeremycorbyn  & official account of Jeremy Corbyn\\
        @Conservatives & official account of the Conservative Party\\
        @UKLabour  & official account of the Labour Party\\
        @CCHQPress & official account of the Conservative Party Press Office\\
        @labourpress & official account of the Labour Party Press Office\\
        @theSNP & official account of the Scottish National Party\\
        @TheGreenParty & official account of the Green Party\\
        @ScottishLabour & official account of Scotland’s democratic socialist party\\
        \cmidrule{1-1}
        @BBCNews & official account of BBC News\\
        @BBCPolitics & official account of BBC's political coverage\\
        @itvnews & official account of ITV News\\
        @ITVNewsPolitics & official account of ITV News for political news\\
        \bottomrule
    \end{tabular}
    \label{tab:false_positives}
\end{table}

In this work, we use a publicly available \cite{nizzoli2021_TwitterDatasetCoordinated} large collection of English tweets gathered in the run-up to the 2019 UK general election \cite{nizzoli2021coordinated}, from 12 November 2019 to 12 December 2019 (election day), and provided to us in full by its original authors. Although having a restricted time frame can generally be a limiting factor in studies that aim to assess social dynamics online, the specific political context that led to the 2019 UK general election guarantees that our dataset includes the vast majority of the debate surrounding the election. In fact, the date of the election was set on October 31 and the campaigning only officially started on November 6, in correspondence with the dissolution of the Parliament. Within this time frame, all tweets that featured at least one of the predefined election-related hashtags in Tab. \ref{tab:data_hashtags} were collected; some of the hashtags have a clear political alignment (Labour or Conservative), while the rest only refer to the election itself and can be considered neutral. Additionally, the dataset includes all tweets published by the official accounts of the two parties and their leaders and all retweets and replies they received, as summarized in Tab. \ref{tab:data_accounts}. The final dataset combines these two collection processes, resulting in a set of 11,264,280 tweets posted by 1,179,659 distinct users.

To estimate the toxicity conveyed by a tweet, we leverage the publicly available Perspective API~\cite{perspectiveAPI}, a state-of-the-art model that is currently an established standard for automatic toxic speech detection. We will thus follow the definition of ``toxic content'' it is based upon, which is ``a rude, disrespectful, or unreasonable comment that is likely to make people leave a discussion''. To achieve more robust results, we standardize all tweets in our dataset by enforcing UTF-8 encoding, converting them to lower-case, and stripping them of all hashtags, URLs, mentions, and emojis. Therefore, we send an API request to Perspective API and obtain within the response the \texttt{TOXICITY} score output by the model for each tweet, a value in the range $[0,1]$ that represents ``the likelihood that someone will perceive the text as toxic'' \cite{perspectiveAPI}. Indeed, Perspective is trained on millions of comments that are annotated with the fraction of human raters who have tagged the given comment as toxic. We note that all tweets that after the aforementioned preprocessing steps resulted in an empty message were not assigned a toxicity score, whereas the small proportion ($<0.5 \%$) of non-English tweets present in the dataset whose language is supported by Perspective API were annotated. Although multiple dimensions of toxicity are provided by Perspective API, similarly to previous works \cite{saveski2021toxic, Avalle2024} we only focus on the \texttt{TOXICITY} score, which is described in Perspective API's official documentation as ``Perspective's main attribute''. Further analyses we have conducted show that the \texttt{TOXICITY} score is highly correlated to the rest (see SI, Sec. 2.1), meaning that it can capture whether a tweet can generally be perceived as toxic or non-toxic more reliably than the other (more specific) attributes.

\subsection{Coordinated communities} 
In recent years, many methodologies to detect coordinated behavior on online social networks have been proposed. Typically, these methods rely on the assumption that highly synchronized activity between users is abnormal and that users exhibiting suspiciously similar behavior -- whether human-controlled or automated -- are likely to be coordinated. Pacheco \textit{et al.} (2021) \cite{pacheco2021uncovering} argue that coordination detection should start from establishing what constitutes suspicious behavior; in this regard, they define several \textit{behavioral traces}, such as activity timestamps, sequences of hashtags, or account handle sharing, and create bipartite networks where users are connected to such traces. Projecting this network results in a network of accounts whose edges may indicate coordination. Luceri \textit{et al.} (2024) \cite{luceri2024unmasking} expand on this framework by introducing the concept of a \textit{fused network} that combines the similarity networks obtained by analyzing multiple behavioral traces. Unlike previous coordination detection techniques, they also rely on node centrality measures and node pruning, arguing that focusing solely on edge weights may fail to capture behavioral similarities. Rather than evaluating specific behavioral traces, other works in the literature focus on encoding user activity. For instance, Cresci \textit{et al.} (2016) \cite{cresci2016dnainspired} leverage DNA-inspired modeling to characterize users and discriminate genuine accounts and spambots, whereas Chavoshi \textit{et al.} (2016) \cite{chavoshi2016debot} employ a novel technique called \textit{warped correlation} to measure the similarity between the activity signals of users.
In their work, Nwala \textit{et al.} (2023) \cite{nwala2023language} report very promising results obtained by their proposed BLOC framework, a platform-agnostic method that encodes activity using an \textit{action alphabet} and a \textit{content alphabet}, which characterizes the lexical content of a post. Using the generated BLOC representations, they proceed by clustering accounts with highly similar feature vectors and study each cluster by the variety of the behavior of the users therein and their extent of automation.

\begin{figure*}[t!]
    \centering
    \includegraphics[width=0.9\textwidth]{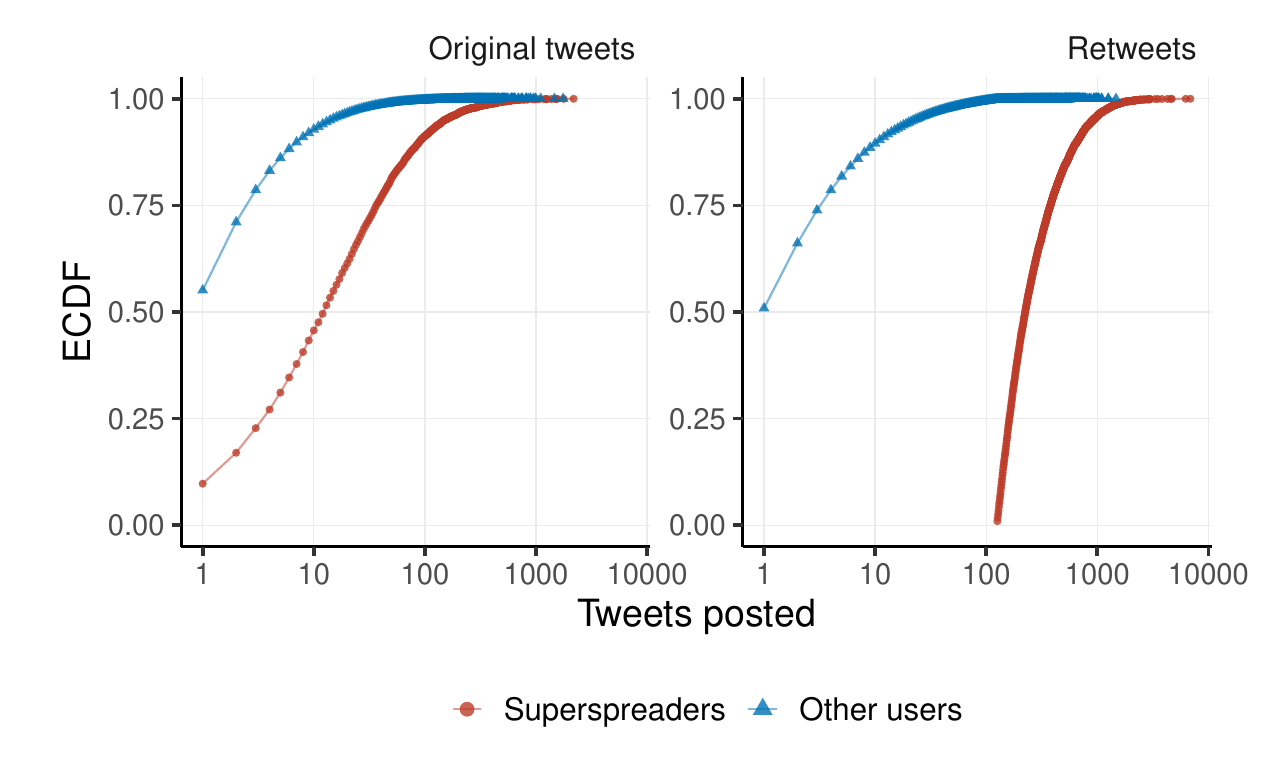}
    \caption{Empirical Cumulative Distribution Function (ECDF) of the number of tweets posted by superspreaders compared to other users, taking into account original tweets (left panel) and retweets (right panel).} 
    \label{fig:tweets_cdf}
\end{figure*}

In this work, we apply the state-of-the-art method proposed by Nizzoli \textit{et al.} (2021) \cite{nizzoli2021coordinated}. The framework they present, which is designed to capture retweet-based coordination, is capable of measuring the extent of coordination of a user on a continuous scale (from ``weakly-coordinated'' to ``strongly-coordinated'') rather than binary (``coordinated'' or ``non-coordinated''), which allows for a finer characterization of coordinated communities. We now detail how their coordination detection algorithm functions and how it was applied to our dataset, which is the same that Nizzoli \textit{et al.} used to showcase their methodology. Their procedure first involves the selection of a set of accounts to be investigated; in our case, we focus on the top 1\% retweeters in the dataset. This operation results in a set of 10,782 accounts which from this point onward we will refer to as \textit{superspreaders}. Despite being characterized by an extremely prolific retweeting activity, in Fig. \ref{fig:tweets_cdf} we show that superspreaders also tend to produce more original tweets than other users. Secondly, the TF-IDF vectors of the IDs of tweets retweeted by each superspreader are computed, resulting in a feature vector for each user. This weighting places more importance on the action of retweeting of unpopular tweets, which is a telltale sign of suspicious behavior. By computing the pairwise cosine similarities between feature vectors, a weighted undirected similarity network is thus obtained. In order to retain only statistically relevant connections, the user similarity network is filtered to extract its multiscale backbone \cite{serrano2009_ExtractingMultiscaleBackbone}, resulting in a network of 10,782 nodes (i.e., the superspreaders) and 276,775 edges. By studying this similarity network, the extent of coordination of each user is then estimated as a measure of how strictly the network must be filtered to disconnect that user. In detail, a moving threshold that is progressively incremented is used to remove edges with a weight lower than the threshold. At each time step, the nodes that have been disconnected are assigned a \textit{coordination score} in $[0,1]$ equal to the percentile rank of the threshold used with respect to the edge weight distribution of the non-filtered network (i.e., the similarity network prior to the application of the moving threshold). These steps are repeated until there are no more edges with weight greater than the current threshold, and therefore all users have been assigned a coordination score.

Using the Louvain clustering algorithm \cite{blondel2008fast}, Nizzoli \textit{et al.} detected a total of seven clusters of coordinated users, which they subsequently characterized politically by assessing the key debates and issues discussed by the users in each community, using TF-IDF weighting of hashtags and word shift graphs. In Sec. \ref{sec:results_tox_coord_comm}, we will focus on providing a toxicity-based characterization for the three largest communities they identified: the ``LAB'' and ``CON'' clusters, populated by users supporting the Labour and Conservative parties, respectively, and the ``TVT'' cluster, a community tightly related to the ``LAB'' cluster that promotes liberal democrats, tactical voting, anti-Brexit, and anti-Tory campaigns. Using the political leaning inference method we detail in Sec. \ref{sec:political_leaning}, we have further validated the rationale behind the names Nizzoli \textit{et al.} have assigned to these clusters. In addition to being the largest, we argue that these three communities are also the most representative of the dataset, and thus most valuable for investigation: the LAB and CON clusters have distinctly politically slanted narratives, whereas the TVT cluster can serve as a benchmark to validate the interplay between political leaning, toxicity, and coordinated behavior.

For the purposes of this study, we label as \textit{coordinated} all superspreaders having a coordination score above the median of the coordination score empirical distribution, which is approximately equal to 0.8. Out of the 5450 superspreaders above this threshold, we have ignored 12 accounts (8 political accounts and 4 news accounts) that upon manual inspection turned out to be false positives, which we report in Tab. \ref{tab:false_positives}. We reckon that the high coordination score that these accounts exhibit can be explained by other (coordinated) users synchronizing with their activity, or with these accounts being evidently coordinated between themselves. Such a limited fraction of accounts can't affect analyses that involve comparing coordinated and non-coordinated users, however, considering the large number of interactions they generate, they can influence estimates derived from the interactions themselves, including those we present in Sec. \ref{sec:results_exposure}, and inflate the number of interactions attributed to CB. For these reasons, the resulting set of coordinated users includes a total of 5438 accounts; we consider all remaining users -- including the aforementioned 12 accounts and all other superspreaders -- to be non-coordinated. Overall, coordinated superspreaders appear to retweet more and quote less (SI, Fig. S1), however their hashtag usage is less diverse than other users (SI, Fig. S2). Furthermore, coordinated superspreaders receive more interactions (i.e., quotes, replies, or retweets) than non-coordinated superspreaders and all other non-coordinated users (SI, Fig. S3), and tend to exhibit stronger signs of automation \cite{nizzoli2021coordinated}.

\subsection{Political leaning inference}
\label{sec:political_leaning}
In this work, we also aim to investigate the role of political leaning and its relationship with coordination and toxicity. The alignment of a user on the political spectrum can be estimated by studying hashtag usage~\cite{martinez2011disentangling, bovet2018validation}. Specifically, we build a tweet-hashtag bipartite network starting from the sets of all tweets in our dataset and the hashtags in them. Then, we project this network onto a hashtag co-occurrence network, whose nodes represent hashtags and an edge between two hashtags indicates the two have appeared at least once in the same tweet; the weight of the edge is a positive integer representing the number of such co-occurrences. The co-occurrence network resulting from this projection contains a total of 100,461 nodes and 822,420 edges. This network can then be used to infer the leaning of each hashtag by applying a label propagation algorithm, using the set of hashtags we defined for the data collection step as an initial seed of known polarity, in a way similar to~\cite{bovet2019influence}.
The algorithm we propose builds upon the multiscale backbone extraction method~\cite{serrano2009_ExtractingMultiscaleBackbone} to identify relevant connections (i.e., co-occurrences) among the network's hashtags. All nodes are initially assigned an ``undefined'' political leaning score, except for the initial seed of hashtags of known polarity used for data collection (Tab. \ref{tab:data_hashtags}). A single iteration $i$ of the algorithm consists of two main operations: extraction of the backbone of the co-occurrence network with a disparity filter $\alpha_i$; and simultaneous update of all hashtags with undefined leaning at step $i$. The leaning assigned to a hashtag is equal to the average of the leaning of its neighbors, weighted by the number of co-occurrences with each. In this computation, all neighbors that weren't already assigned a leaning in a previous step are temporarily assigned a leaning of 0. At step $i+1$, a larger disparity filter $\alpha_{i+1} > \alpha_i$ is used to extract the backbone; this corresponds to a softer filtering that allows more nodes to be part of the extracted network. The newly added hashtags and those that haven't been labeled in the previous steps will thus be assigned a political leaning score. The algorithm stops when all hashtags have been assigned a score or when a disparity filter equal to 1 is employed, meaning the update is performed on the entire network. In the latter case, all hashtags that still haven't been updated at the end of the algorithm will be assumed to be neutral and assigned a political leaning score of zero. The length of the sequence of disparity filters will have an impact on the final result of the label propagation, as it's strongly dependent on the network it is applied on and on the nodes used as the initial seed: defining an excessively short sequence may lead to many neutral hashtags, while a long one -- although preferable -- can become computationally expensive if the network involved is large. For our purposes, we have found choosing a sequence of values that scale logarithmically and span across several orders of magnitude to be the optimal choice, in contrast to setting a fixed increment to be added at each iteration.

Knowing the polarity of hashtags allows us to get an estimate of the political leaning of each tweet, which we define as equal to that of the hashtag in it with the highest score in absolute value (i.e., most polarized); this avoids the leaning of polarized hashtags being averaged out in tweets where they are used in conjunction with many politically neutral hashtags. Finally, we estimate the political leaning of a user as the average political leaning score of its tweets and retweets.

\section{Results and Discussion}
\label{sec:results}

\begin{figure*}[t!]
    \centering
    \includegraphics[width=1\textwidth]{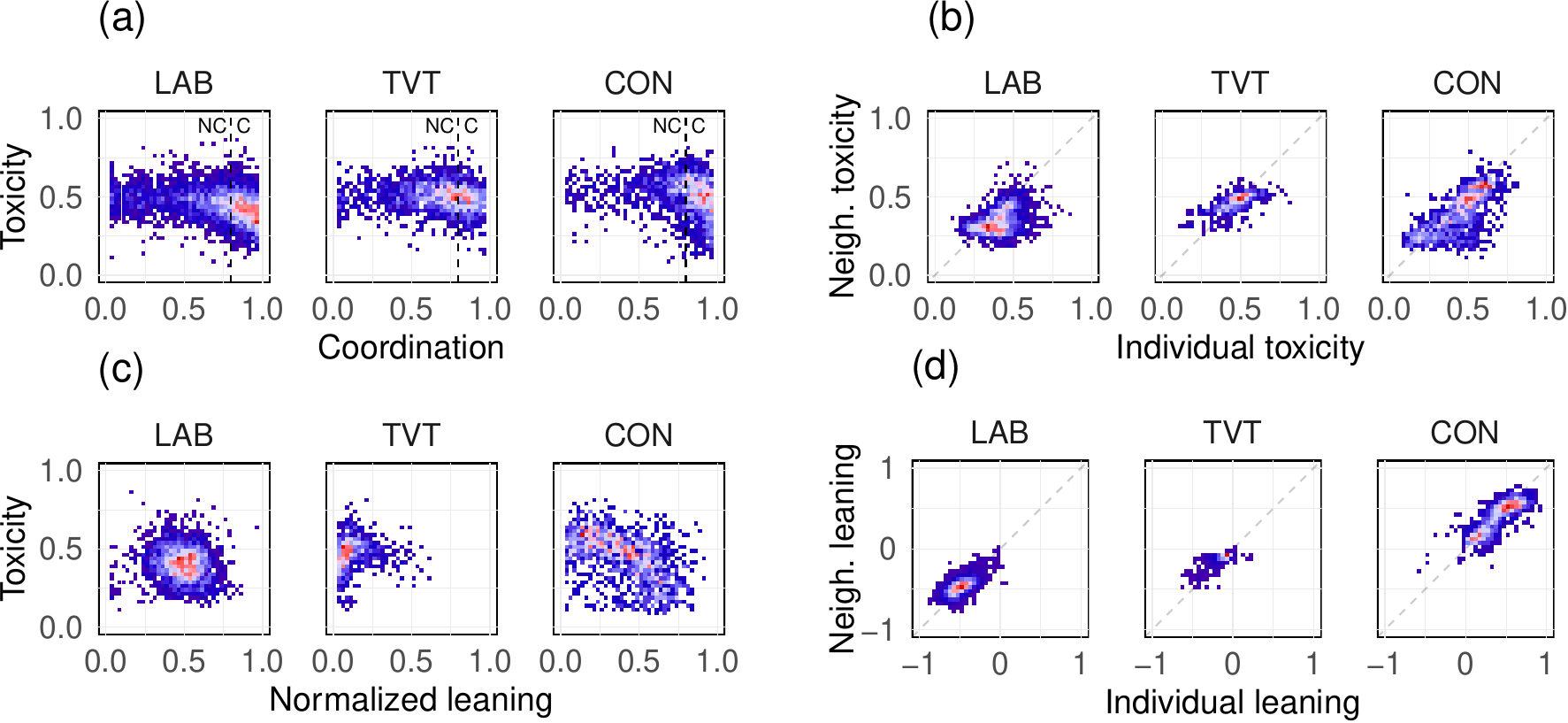}
    \caption{Joint distributions of different metrics for the superspreaders in the three largest communities of the similarity network. Color intensity indicates the number of users in a bin, with red regions highlighting peaks. (a) Coordination score and toxicity of users, with the dashed line (median coordination) indicating the threshold used to label a superspreader as ``coordinated''; (b) toxicity of users and weighted average of their neighborhood's toxicity; (c) cluster-normalized (least to most extreme) political leaning of users and their toxicity; (d) political leaning of users and weighted average of their neighborhood's leaning.} 
    \label{fig:rq1_grid}
\end{figure*}

\subsection{Toxicity across coordinated communities}
\label{sec:results_tox_coord_comm}
To analyze variations in toxicity across coordinated communities, each user is assigned a toxicity score. This score is determined by averaging the toxicity of a user's top 10\%  most toxic original tweets and retweets. By focusing on this metric, we aim to gauge the peak toxicity a user can manifest and disseminate~\cite{cinelli2021dynamics, cheng2017anyone} rather than their standard activity levels. Indeed, although ``pure'' haters and highly toxic comments are rare, their presence can still have a detrimental impact on discourse \cite{cinelli2021dynamics}, and using a metric that is more sensitive to ``extreme'' behavior can provide a more meaningful characterization of users. In particular, the rationale for choosing the ``top 10\%" over other thresholds was to provide an interpretable definition of user toxicity, according to which most users exhibit ``average" toxic behavior and only a smaller number sit in the right (``highly toxic") and left (``rarely toxic") tails, while making sure that the score for each user is computed by averaging the toxicity of a large enough sample of tweets. More details on this aspect of our methodology are provided in Sec. 2.1 of the Supplementary Information.

Fig. \ref{fig:rq1_grid}(a) displays the joint Probability Density Function (PDF) of the coordination scores of superspreaders from the three largest clusters within the similarity network, aligned with their toxicity (as defined previously). A discernible trend emerges: strongly-coordinated users tend to display lower toxicity. This pattern, which is particularly pronounced within the Labour and Conservative clusters (i.e., the most distinctly politically aligned communities of superspreaders) implies a propensity among coordinated users to disseminate content with low toxicity levels. We can validate the observed relationship between coordination and user toxicity by computing Spearman's correlation coefficient $\rho$ for the coordinated and non-coordinated superspreaders in each cluster. For the former, we obtain $\rho=-0.34$ (95\% CI: $[-0.37, -0.30]$) for the LAB cluster, $\rho=-0.21$ (95\% CI: $[-0.28, -0.14]$) for the TVT cluster, and $\rho=-0.46$ (95\% CI: $[-0.50, -0.42]$) for the CON cluster; for the latter, we obtain $\rho=-0.18$ (95\% CI: $[-0.21, -0.14]$) for the LAB cluster, $\rho=-0.035$ (95\% CI: $[-0.085, 0.017]$) for the TVT cluster, and $\rho=-0.020$ (95\% CI: $[-0.090, 0.051]$) for the CON cluster. Furthermore, Fig. \ref{fig:rq1_grid}(a) highlights the coordination score threshold employed to classify a superspreader as `coordinated'; subsequent analyses will center on this specific user subset.

Fig. \ref{fig:rq1_grid}(b) depicts the joint PDF of each coordinated user's toxicity against the average toxicity of their neighboring users. These neighbors are weighted by their similarity (i.e., edge weight) to the user. The displayed PDF exposes observable toxicity-based homophily in the CON cluster and, to a lesser extent, in the TVT cluster, suggesting that users with comparable retweet behaviors within these clusters also exhibit similar toxicity levels.
To confirm this intuition, we can consider the subgraph corresponding to each cluster and estimate its assortativity~\cite{newman2003mixing} with respect to the toxicity of the users within, resulting in a score of 0.53 for the CON cluster, 0.30 for the TVT cluster, and 0.06 for the LAB cluster. Shuffling the toxicity scores among users of the same cluster allows us to validate these estimates further \cite{anagnostopoulos2008Influence}; after 10,000 random shuffles, we obtain an average assortativity coefficient approximately equal to zero and a Z-score $\gg 1$ for the observed coefficients of the CON and the TVT clusters.

In Fig. \ref{fig:rq1_grid}(c), we present the joint distribution of user toxicity juxtaposed with a normalized political leaning score (see Fig. S11 in SI for a comparison of toxicity and political leaning involving all users). Interestingly, the distribution pertaining to the coordinated users in the CON cluster suggests that the toxicity they express and their political alignment are negatively correlated; in fact, for the two metrics we obtain a Spearman's correlation coefficient of $\rho = -0.50$ (95\% CI: $[-0.54, -0.46]$).
For additional context, Fig. \ref{fig:rq1_grid}(d) illustrates the relationship between the political leaning of coordinated users and their neighbors. The distributions for the LAB and TVT clusters are mostly centered around a singular peak; this is especially evident in the latter, where users predominantly employ non-polarized hashtags and thus form a distinct peak around zero. On the other hand, the distribution of the CON cluster is characterized by two separate peaks, one close to 0.5 -- similarly to the LAB cluster, although with opposite alignment -- and the other close to zero. This result suggests that the CON cluster might effectively house at least two sub-communities, as denoted by the two observed peaks. The peak around 0.5 may be comprised of coordinated users sharing tweets with a distinct political connotation, which is determined by hashtags often co-occurring with the conservative hashtags of polarity $(+1)$ used for data collection (Tab. \ref{tab:data_hashtags}). On the other hand, the coordinated users distributed around zero may still be sharing tweets with a clear conservative perspective, however, their focus might be on topics characterized by hashtags that are not tightly linked to our predefined set of conservative hashtags. This observed within-cluster bi-modality is further strengthened by our previous analysis on the relationship between polarity and toxicity: the users in these potential sub-communities not only share tweets with different political content, as suggested by their hashtag usage, but also with different linguistic content, as suggested by the toxicity they express. Specifically, users sharing more distinctly politically aligned tweets do so with seemingly non-toxic language. This result further highlights the nuanced behavior of coordinated users in regard to the topics they promote and how they promote them from a linguistic viewpoint.

\subsection{Toxicity in coordinated and non-coordinated users}
\label{sec:results_coord_vs_noncoord}

\begin{figure*}[t!]
    \centering
    \includegraphics[width=1\textwidth]{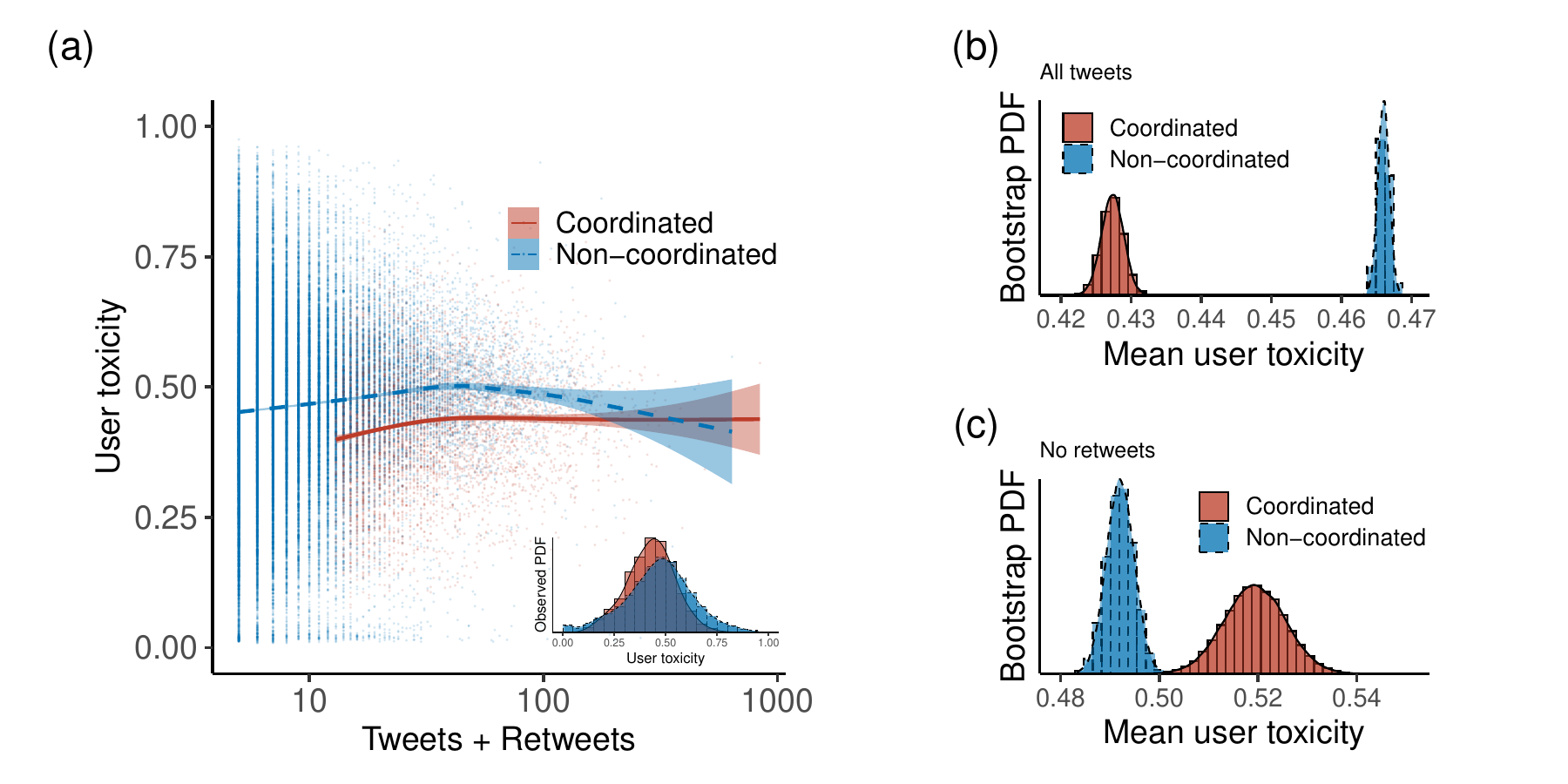}
    \caption{Comparison between coordinated and non-coordinated users in terms of expressed toxicity, defined as the average of the top 10\% most toxic original tweets or retweets. (a) Tweeting activity and user toxicity smoothed via a GAM fit (the shaded region indicates the corresponding 95\% CI), with the observed user toxicity distribution for both groups in miniature; (b) bootstrap distribution of the average user toxicity; (c) bootstrap distribution of the average user toxicity obtained by ignoring retweets.} 
    \label{fig:rq2_grid}
\end{figure*}

In this analysis, we seek to measure the difference in toxicity exhibited by coordinated and non-coordinated users, applying for both the ``user toxicity'' definition we presented in Sec. \ref{sec:results_tox_coord_comm}. For what concerns this comparison, we only consider users with a minimum of 5 tweets or retweets, as including all users would generate a user toxicity distribution with a heavy positive skew for both groups, thus making the comparison less valuable for an assessment of the overall behavioral differences. 

Fig. \ref{fig:rq2_grid}(a) shows that the toxicity expressed by coordinated users remains relatively unchanged across different activity levels, suggesting that the more active coordinated users don't share content that is any more or less toxic. Examining the user toxicity distributions for both user groups in Fig. \ref{fig:rq2_grid}(a) highlights that coordinated users display a distribution that is more sharply concentrated around its mean ($\overline x=0.43, \hat\sigma=0.12$), while non-coordinated users are characterized by larger fluctuations ($\overline x=0.47, \hat\sigma=0.17$). We have statistically validated ($p<0.001$) that the two samples don't originate from the same population by means of the Anderson Darling $k$-sample test with 10,000 simulations. We also note that these results are robust to the specific threshold used to define ``user toxicity'' (see SI, Sec. 2.2). Our findings provide an indication that coordinated users manifest on average lower toxicity than their non-coordinated counterparts.  We can quantify this discrepancy in average toxicity via bootstrap resampling, to account for the inherent difference in sample size between the two groups. In Fig. \ref{fig:rq2_grid}(b), we report the distributions resulting from 50,000 bootstrap replicates, which further show that coordinated users are on average less toxic than the non-coordinated ones; for the former, we obtain $\hat\mu = 0.4273$ (95\% CI: $[0.4243, 0.4305]$), while for the latter
$\hat\mu = 0.46599$ (95\% CI: $[0.46439, 0.46759]$). We argue that this result follows from the difference in tweeting activity of the two groups. In fact, we have identified coordinated users as a subset of the most prolific retweeters in the dataset, hence their activity being mostly characterized by sharing other accounts' content suggests that coordinated behaviors may favor promoting content that is less toxic than average. To assess whether this result effectively stems from the difference in tweeting patterns, we repeat the same analysis upon exclusion of retweets, thus solely focusing on coordinated users who also post original tweets: out of the 5438 users we initially identified, this filtering reduces the set of coordinated users to 4503. The densities obtained via bootstrap reported in Fig. \ref{fig:rq2_grid}(c) indicate that extensive retweeting activity does play a role, and also elicit an interesting observation: both coordinated and non-coordinated users tend to post original content with a higher toxicity level compared to that they disseminate through retweets. In fact, we measure $\hat\mu = 0.5193$ (95\% CI: $[0.5072, 0.5315]$) for coordinated users, and $\hat\mu = 0.4921$ (95\% CI: $[0.4866, 0.4976]$) for the non-coordinated.
In addition, our analysis highlights that coordinated users produce original tweets that are distinctly more toxic than those they spread via retweeting, whereas non-coordinated users tend to produce and retweet content of similar toxicity. Indeed, unlike our previous analysis that included retweets, the average toxicity of the coordinated users is now higher than that estimated for the non-coordinated. This suggests that coordinated behaviors, aiming to maximize exposure, might be strategically orchestrated to disseminate toxic content while remaining wary of triggering platform moderation mechanisms.

\subsection{Interactions with coordinated users and toxic production}
\label{sec:results_exposure}

\begin{figure*}[t!]
    \centering
    \includegraphics[width=1\textwidth]{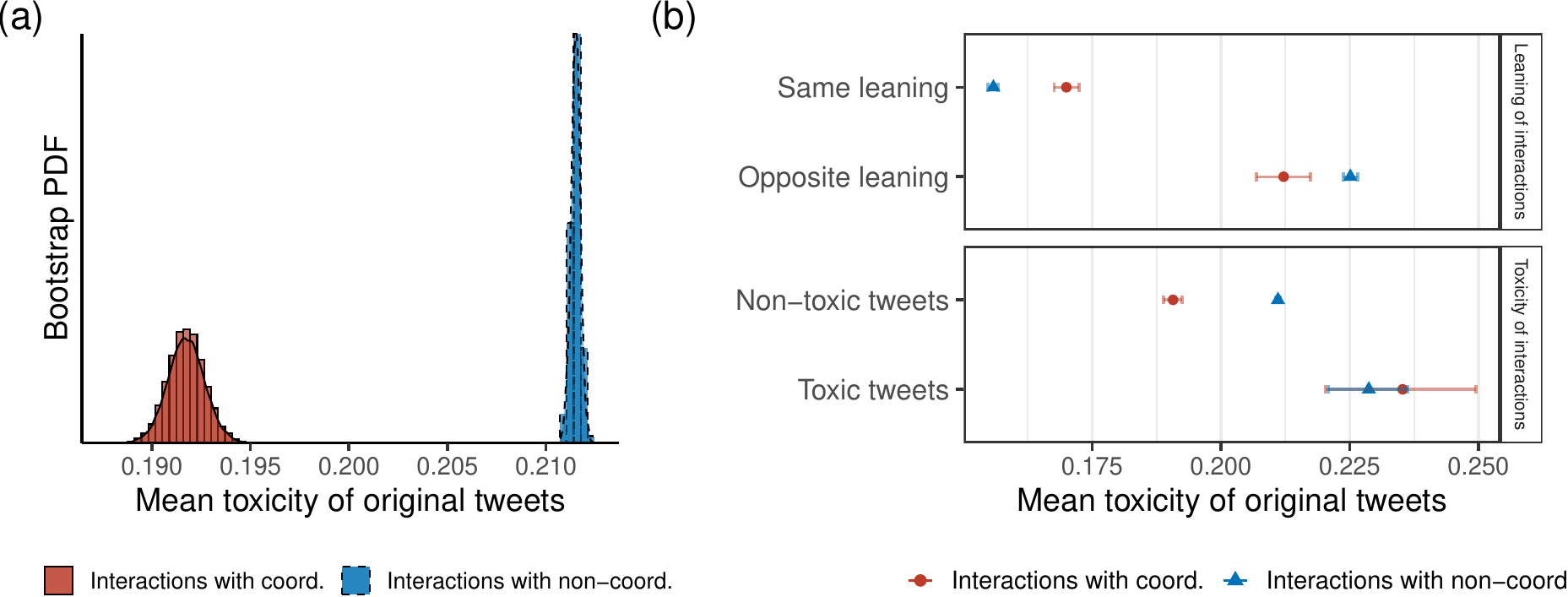}
    \caption{Average toxicity of tweets produced by non-coordinated users, obtained with bootstrap resampling. (a) Distributions of the average toxicity produced following exclusive interactions with non-coordinated users or coordinated users; (b) estimates with their 95\% CI obtained by factoring in the toxicity of the interactions, using a score of 0.6 as the threshold to label a tweet as `toxic', and their political leaning.} 
    \label{fig:rq3_grid}
\end{figure*}

We further examine whether interacting with coordinated users corresponds to an increased production of toxic content by the non-coordinated. To achieve this, we classify the tweets each non-coordinated user has been involved in into two main categories: `productions' (i.e., the original tweets the user has written) and `interactions' (i.e., the tweets the user has interacted with). Similarly to previous works \cite{cresci2016dnainspired, nwala2023language}, this classification allows us to encode user activity within the context of the electoral debate as a sequence of productions and interactions that we can study from a toxicity viewpoint. For instance, let us consider a user who in order performs the following actions: post a simple tweet, retweet two tweets, post two simple tweets, reply to a tweet, quote a tweet, retweet a tweet. We encode this sequence as \texttt{T>RT>RT>T>T>[R]>R>[Q]>Q>RT}, where ``\texttt{T}'' stands for ``simple tweet'', ``\texttt{RT}'' for ``retweet'', ``\texttt{R}'' for ``reply'', ``\texttt{Q}'' for ``quote'', $[\cdot]$ for the tweet that is referenced (e.g., the tweet the user has replied to), and ``\texttt{>}'' for the chronological order of the sequence.
This ordered sequence can thus be encoded as \texttt{P>I>I>P>P>I>P>I>P>I} where ``\texttt{I}'' stands for interaction and ``\texttt{P}'' for production. As we’re interested in studying the potential relationship between the toxicity of the content the user interacts with and that of the content the user later produces, we remove the first production (or sequence of productions) if not preceded by one or multiple interactions, and the last interaction (or sequence of interactions) if not followed by one or multiple productions. In the previous example, the final user sequence is thus \texttt{I>I>P>P>I>P>I>P}. Finally, we characterize each block of consecutive interactions by measuring the fraction of ``toxic tweets'', the fraction of coordinated users involved, and the average political leaning, and each block of consecutive productions by measuring its average toxicity. To assess the role of CB, we only focus on sequences of interactions that involve only coordinated or non-coordinated users. This approach allows us to obtain for each user a series of toxicity scores corresponding to the content the user has produced and the kind of interactions the user had preceding each of those. Finally, we concatenate the data of all users and measure the average toxicity produced after different kinds of interaction, which we now detail.

In Fig. \ref{fig:rq3_grid}(a), we report the density of the average toxicity produced by non-coordinated users, obtained via bootstrap resampling, upon interacting with both groups of users. The resulting estimates suggest that non-coordinated users who have exclusively interacted with coordinated behavior tend to manifest slightly lower toxicity levels than those who haven't at all; we obtain $\hat\mu = 0.19173$ (95\% CI: $[0.18987, 0.19356]$) for the former, and $\hat\mu = 0.21150$ (95\% CI: $[0.21102, 0.21198]$) for the latter. 

Fig. \ref{fig:rq3_grid}(b) displays how factoring in the toxicity of the content a user has interacted with affects the previous outcome. In this analysis, we consider a sequence of interactions to be `toxic' if all the tweets that constitute it have a toxicity score above 0.6; in contrast, a sequence is considered `non-toxic' if all of its tweets have a toxicity score below 0.6. The choice of the threshold is motivated by how Perspective was trained to classify a tweet as toxic: a threshold of 0.6 indicates with reasonable confidence that more than half the readers would classify that message as such. However, further experiments we have conducted indicate that the results of this analysis are robust to the specific threshold employed (see SI, Sec. 2.2). Finally, we discard all sequences with a mix of toxic and non-toxic tweets. Our results indicate that the level of toxicity a user expresses upon interacting with toxic content only slightly changes when the author of that content is a coordinated user. In fact, the two bootstrap estimates largely overlap: $\hat\mu = 0.2353$ (95\% CI: $[0.2203, 0.2495]$) following interactions with coordinated users, and $\hat\mu = 0.2287$ (95\% CI: $[0.2209, 0.2363]$) with non-coordinated users. In the case of interactions with non-toxic tweets, we obtain that interacting with coordinated users doesn't correspond to a more toxic production of original tweets: $\hat\mu = 0.19071$ (95\% CI: $[0.18888, 0.19247]$) following interactions with coordinated users, and $\hat\mu = 0.21109$ (95\% CI: $[0.21060, 0.21157]$) with non-coordinated users. 
To investigate this further, we consider the political leaning of the content a user has interacted with and compare it to the user's own leaning: if the average leaning of the interactions has an opposite sign to that of the user, we consider that sequence of interactions to have `opposite leaning', otherwise `same leaning'. To avoid ties, we exclude from this analysis all users with a political leaning score equal to 0. Fig. \ref{fig:rq3_grid}(b) shows that users interacting with content of opposite political leaning exhibit slightly higher toxicity levels than those interacting with politically aligned content. In addition, these findings point towards coordinated behavior having only a minor role in affecting the toxicity of tweets produced by non-coordinated users: the toxicity level and the political content of the tweets being shared have a similar or larger impact than the coordination degree of its author. In fact, our measurements suggest that the `volume' of tweets, which is characteristic of coordinated behaviors, has a less pronounced impact on the toxic behavior of non-coordinated users than the political message they convey and how they convey it from a toxicity viewpoint.

\begin{figure*}[t!]
    \centering
    \includegraphics[width=1\textwidth]{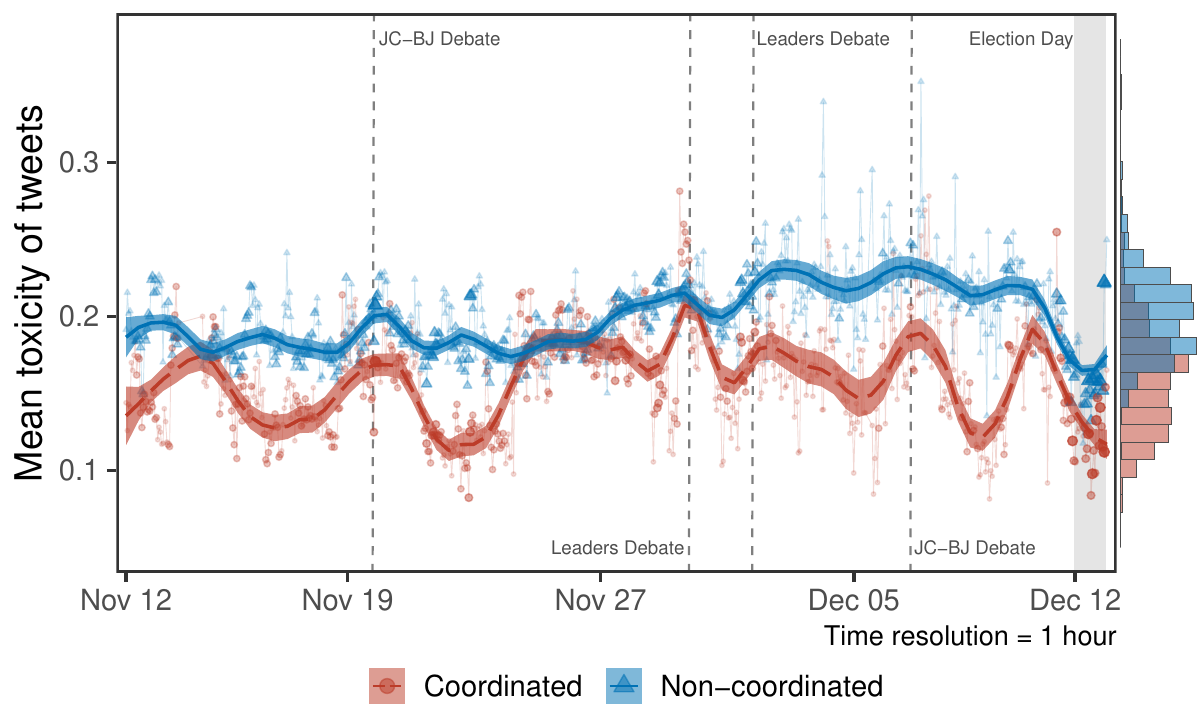}
    \caption{Hourly average toxicity of the tweets produced by non-coordinated users compared with that of the tweets by coordinated users they have interacted with, overlaid with LOESS curves (the shaded regions indicate their respective 95\% CI) for easier visualization. The size and visibility of a point are proportional to the number of tweets observed within the corresponding hour. On the right side, empirical distributions of the two metrics across the entire time period.} 
    \label{fig:rq3_transfer_entropy}
\end{figure*}

An alternative approach we can employ to explore how CB might potentially relate to toxic production is to study the phenomenon from a strictly temporal perspective. In this regard, we construct two time series spanning the entire time frame of our data: one is built by sequencing the hourly average toxicity of original tweets produced by non-coordinated users, while the other by averaging the toxicity of the tweets they have interacted with that have been shared by coordinated users. Fig. \ref{fig:rq3_transfer_entropy} shows that while the former appears stable and concentrated around a small range of values, the latter is characterized by evident oscillations. This hints at coordinated users adopting a noticeably more toxic tone in the context of specific events in the campaign (such as TV debates between political leaders) and less toxic in others. As we have suggested in Sec. \ref{sec:results_coord_vs_noncoord}, a possible explanation for this behavior might reside in CB being specifically employed to inject inflammatory content or trigger toxic responses while simultaneously avoiding the platform moderation mechanism.

Furthermore, from Fig. \ref{fig:rq3_transfer_entropy} we can observe that the two-time series nearly mirror each other in proximity to election day, which suggests the two might be related. To verify this, we can quantitatively measure the information flow between the two by applying the transfer entropy method \cite{schreiber2000Measuring}, which is based on the concept of entropy. Unlike other methods such as Granger causality \cite{granger1969Investigating}, transfer entropy can capture not only non-linear dependencies between two time series, but also the dominant direction of the flow, by estimating the net gain in information about the future observations of one derived from the past observations of the other. As entropy requires discrete data, we discretize the average toxicity scores of the two-time series and assign them to 4 bins with bounds equal to the quantiles at $\mathbf p = (0.05, 0.5, 0.95)$ of their respective empirical distributions. The selection of these quantiles is made to emphasize the tails of the distributions, as they include the least and most toxic observations; to this end, we compute the Rényi transfer entropy with a weighting parameter of $q = 0.5$, which puts more weights on the tails when calculating transfer entropy \cite{behrendt2019RTransferEntropy}. Our findings indicate that the information flow from the toxicity of content shared via CB and that produced by non-coordinated users is not statistically significant ($p=0.31$), as tested using the method proposed by Dimpfl and Peter (2013) \cite{dimpfl2013UsingTransferEntropy} and implemented by Behrendt \textit{et al.} (2019) \cite{behrendt2019RTransferEntropy}; the same considerations apply for the information flow in the opposite direction ($p=0.67$). Interestingly, repeating the same analysis by not restricting to interactions with coordinated users yields a different result: the information flow from the toxicity of interactions to the toxicity of produced content is statistically significant ($p=0.0051$), whereas that in the opposite direction is still not ($p=0.38$). This suggests that while the toxicity produced by non-coordinated users may overall be correlated to that they have interacted with, this relationship can't be solely ascribed to the activity of coordinated users. 

Our findings expose that interacting with CB doesn't correspond to a more toxic production by non-coordinated users. This result is likely attributed to the nature of the content being shared, suggesting that the `character' of the message (such as its political slant) might play a bigger role than the extent of coordination of the efforts that channeled it. Additionally, our results are somewhat counterintuitive concerning the current assumption regarding political polarization.

\section{Limitations}
While our work offers novel insights into the interplay between coordinated behavior and toxic content dissemination, we acknowledge that it also presents some limitations that should be addressed. Regarding measuring the toxicity of tweets, we relied on Perspective API because it is one of the most advanced models for toxicity detection, and the use of automated tools is unavoidable for annotating large datasets. However, studies have shown that models such as Perspective API may also suffer from several limitations, including bias, reproducibility issues, or mishandling the multifaceted nature of abusive language \cite{Vidgen2020, rosenblatt-etal-2022-critical, sheth2022, Pozzobon2023}. For what concerns the detection of coordinated users, we employed a method \cite{nizzoli2021coordinated} that is based on selecting a subset of highly active users (i.e., the top 1\% retweeters) and subsequently studying their similarity in terms of their activity by clustering together highly similar users. While this method has been shown to yield well-separated communities of superspreaders, one of its drawbacks is that it might not capture the full extent of coordinated actions observed online and thus hinder the heterogeneity of the network. For instance, some actors may exhibit coordination that is not retweet-based, or might only work in certain peak periods rather than the whole time frame. In this regard, future studies could provide a more comprehensive view of retweet-based coordinated behavior by analyzing the entire set of users in the data, rather than focusing on a subset, and by then employing appropriate filtering techniques such as the disparity filter \cite{serrano2009_ExtractingMultiscaleBackbone} or the Pólya urn filter \cite{Marcaccioli2019}. However, the computational cost of pairwise comparisons between users may render this kind of approach infeasible for large datasets.

\section{Conclusions}
In the context of the 2019 UK general election, we systematically analyze the interplay between Coordinated Behavior (CB) and content toxicity on X (formerly Twitter). Drawing from a dataset of 11 million tweets from 1 million unique users, we aim to understand better the subtleties of this relationship on a major platform for digital discourse. Our findings indicate that strongly-coordinated users predominantly share content with lower toxicity levels.
This trend is particularly visible in politically affiliated clusters, indicating that coordination, while strategic, does not necessarily result in promoting harmful content. The primary aim of these coordinated users is amplification and influence rather than explicit toxicity dissemination. One potential explanation behind this might reside in coordinated users trying to avoid incurring into the platform's moderation mechanisms, as an extensive resharing of toxic or even hateful content would jeopardize both the information operation they might be driving and the accounts whose tweets they are actively retweeting. A further indication of this is provided by the wavy behavior of the toxicity shared via CB over time, which displays peaks almost exclusively in correspondence with key events that occurred during the campaign, namely the TV debates between the most prominent political leaders involved.

Analyzing coordinated versus non-coordinated users, we find that both groups exhibit comparable average toxicity levels, regardless of their activity volume. Additionally, while both groups tend to produce original tweets more toxic than those they share, this increase is more noticeable for coordinated users, further indicating that CB might be designed to share toxic content while simultaneously escaping moderation mechanisms.
In examining the activity of non-coordinated users, we find that interactions with CB generally don't correspond to increased toxic output, suggesting that the nature of the content might be more impactful than coordinated activity by itself. Similar conclusions are drawn from studying the phenomenon from a temporal perspective, using Rényi transfer entropy to quantify the mutual information flow between toxic production of non-coordinated users and toxicity of the content they interact with.

In conclusion, our study highlights the nuanced manner in which CB functions online in the sphere of toxic speech. While coordination is not directly linked to the spreading of toxicity, its relationship with user behavior is not trivial and begs further investigation. The interplay between content nature and its dissemination strategy is crucial. Future research should investigate the motivations behind coordinated activities, the socio-political environments nurturing them, and their broader effects on digital conversations.

\section*{Funding}
\label{sec:funding}
\noindent
The work is supported by IRIS Infodemic Coalition (UK government, grant no. 
SCH-00001-3391); SERICS (PE00000014); PRIN 2022 ``MUSMA'' (CUP G53D23002930006) funded by EU - Next-Generation EU – M4 C2 I1.1; ``SoBigData.it – Strengthening the Italian RI for Social Mining and Big Data Analytics'' (IR0000013) under the NRRP MUR program funded by the European Union - NextGenerationEU; project CRESP from the Italian Ministry of Health under the program CCM 2022; project SEED N. SP122184858BEDB3.

\section*{CRediT authorship contribution statement}
\label{sec:contribution_statement}
\noindent
\textbf{Edoardo Loru}: Conceptualization, Formal analysis, Investigation, Methodology, Software, Writing – original draft, Writing – review \& editing, Visualization. \textbf{Matteo Cinelli}: Conceptualization, Methodology, Writing – original draft, Writing – review \& editing. \textbf{Maurizio Tesconi}: Conceptualization, Funding acquisition, Methodology, Supervision, Writing – original draft. \textbf{Walter Quattrociocchi}: Conceptualization, Funding acquisition, Methodology, Supervision, Writing – original draft.

\bibliographystyle{unsrt}

\end{document}